\documentclass[apj]{emulateapj}
\usepackage{natbib}
\usepackage{graphicx}
\usepackage{amsmath}
\usepackage[T1]{fontenc}
\usepackage{epsfig}
\usepackage{epstopdf}

\newcommand{\mean}[1]{\mbox{$\langle#1\rangle$}} 
\newcommand{\lsun}{\mbox{L$_\odot$}}
\newcommand{\msun}{\mbox{M$_\odot$}}
\newcommand{\eten}[1]{\mbox{$10^{#1}$}}

\newcommand\sfryso{\mbox{SFR(YSO count)}}
\newcommand\sfrmir{\mbox{SFR(24 \micron)}}
\newcommand\sfrlir{\mbox{SFR($L_{TIR}$)}}
\newcommand\lir{\mbox{$L_{TIR}$}}
\newcommand{\halpha}{\mbox{H$\alpha$}}
\newcommand{\jj}[2]{\mbox{$J = #1\rightarrow#2$}}

\newcommand{\hii}{\mbox{\ion{H}{2}}}
\newcommand{\kms}{\mbox{km s$^{-1}$}}
\newcommand{\av}{\mbox{$A_{\rm V}$}}

\begin{document}

\title{Testing 24 \micron\ and Infrared Luminosity as Star Formation Tracers for
Galactic Star Forming Regions}

\author{Nalin Vutisalchavakul, Neal J. Evans II}
\affil{The University of Texas at Austin, Department of Astronomy,
2515 Speedway, Stop C1400 Austin, TX 78712-1205, USA
}

\begin{abstract}
We have tested some relations for star formation rates used in extra-galactic
studies for regions within the Galaxy. In nearby molecular clouds, 
where the IMF is not fully-sampled, the dust emission
at 24 \micron\ greatly underestimates star formation rates (by a factor of 100 on average)
when compared to star formation rates determined from counting YSOs. 
The total infrared emission does no better. In contrast, the total far-infrared method agrees
within a factor of 2 on average
with star formation rates based on radio continuum emission for massive,
dense clumps that are forming enough massive stars to have \lir\ exceed
\eten{4.5} \lsun. The total infrared and 24 \micron\ 
also agree well with each other for both nearby, low-mass star forming regions and 
the massive, dense clumps regions. 
\end{abstract}


\section{introduction}

Star formation is a fundamental process in the
formation and evolution of galaxies (Kennicutt 1998b, 
Hopkins 2004, Bigiel et al. 2008, Gao \& Solomon 2004). 
A unified picture of star formation across different scales and 
types of regions would benefit from unified measures of 
star formation rates (Krumholz et al. 2011a, 2011b; 
Schruba et al. 2011; Shi et al. 2011; Kennicutt 1998a). 
The most direct way to measure the rate of star formation is to 
count stars of a known age and mass.  Because most galaxies are
too far away for individual star forming regions to be resolved, 
alternative measures of star formation rates have been developed.

Many different methods have been used to estimate the star
formation rate (SFR) in galaxies (Kennicutt 1998b, hereafter K98). 
Commonly used tracers include continuum UV emission,
recombination lines of hydrogen and other atomic species, total infrared
luminosity (\lir ), monochromatic infrared emission, and radio emission (Kennicutt
1998b; Kennicutt et al. 2003, 2009; Calzetti et al. 2007, 2010; Perez-Gonzalez 
et al. 2006; Murphy et al. 2011; Kinney et al. 1993; Condon 1992). 
Each of these indicators
traces star formation in somewhat different ways, averaging over
different timescales (e.g., Kennicutt \& Evans 2012).
UV continuum emission in the wavelength range of 125-250 nm directly
measures radiation from high mass stars, with peak contributions from
stars of several \msun; consequently, it can average SFR over 10-200 Myr.
Hydrogen recombination lines, such as H$\alpha$, or free-free
radio continuum emission trace \hii\ regions surrounding 
high mass stars ($M > 15 \msun$), with
a peak contribution from $M = 30$ to 40 \msun; thus they average SFR over
only 3-10 Myr (Kennicutt \& Evans 2012 and references therein).  

Most studies of star formation in galaxies use UV continuum or optical lines 
(e.g.,  Bigiel 2008, Kinney 1993, Salim 2007, Hao et al. 2011).  
However, optical emission can be strongly affected by dust-extinction,
and the UV continuum is even more sensitive to extinction (Calzetti 1994, 
Hao et al. 2011, Buat et al. 2005, Burgarella et al. 2005). 
The recombination lines trace only very massive stars, so they are sensitive
to assumptions about the IMF (see Figure 1 in Chomiuk \& Povich 2011).

As supplements to UV and
optical tracers, IR fluxes have been used to study SFR in regions that are
obscured by dust (Calzetti et al. 2007, 2010; Perez-Gonzalez et al. 2006; 
Kennicutt et al. 2009).  
Infrared dust emission 
traces the stellar luminosity that has been absorbed by dust
and reemitted in the infrared (K98, Calzetti et al. 2007). It is less biased
towards the highest mass stars and hence less sensitive to the IMF.
If all the photons
inside star forming regions get absorbed by dust, then the total 
infrared emission from dust (\lir) should trace the total luminosity of the 
stars. One problem with using \lir\ to trace star formation is that 
sources other than young stars, such as older stars or AGNs, can 
contribute to heating the dust. For galaxies less active in star formation, a
significant amount of dust heating can come from the general interstellar
radiation field, arising from older stellar populations
(K98, Draine et al. 2007). In that case, \lir\ would trace emission that
is not relevant to the current star formation. 

Monochromatic IR emission has 
also been widely used. One particularly widely used tracer
is the 24 \micron\ continuum emission (Calzetti et al. 2007, Wu et al. 2005a,
 Rieke et al. 2009, Alonso-Herrero et al. 2006, Helou et al. 2004). 
In principle, 24 \micron\ emission has the advantage over \lir\ 
that it requires quite warm dust. In active star forming regions,
the warm dust is more intimately associated with the forming stars.
The diffuse part of the interstellar medium that has been heated by
the average interstellar radiation field should be at a comparatively low
temperature and should not emit much in the 24 \micron\ wavelength band compared
to the emission from high mass star forming regions. 
Stronger radiation fields from
high mass stars can heat the dust to higher temperatures over a larger
region; therefore, 24 \micron\ emission should be a good tracer for 
high mass star forming regions with less contamination from 
non-star-forming sources. 

There are several
studies of how emission from non-star-forming sources compares to emission
relevant to star formation in the 24 \micron\ wavelength (Rahman et al. 2011, Verley
et al. 2008, Draine et al. 2007). Draine et al. showed from fitting dust models to numbers of
galaxies that for galaxies with high star formation rates (starburst galaxies),
the main contribution to the 24 \micron\ emission comes from photodissociation
regions associated with high mass stars. For high mass star forming regions,
24 \micron\ emission should be a good tracer of SFR.
Observations of nearby galaxies
show strong concentrations of 24 \micron\ emission toward \hii\ regions,
but with a diffuse component.

Unifying studies of star formation in other galaxies with studies
within the Milky Way can be mutually illuminating. Chomiuk and
Povich (2011) have compared tracers of SFR on global scales and
found a potential discrepancy of a factor of two between extragalactic
relations applied to the Milky Way as a whole and more direct measures
of the Milky Way star formation rate. Our goal is to test extragalactic
relations on still smaller scales of individual clouds and dense clumps.

Images of the Galactic Plane at 24 \micron\ are available from MIPS
on Spitzer from the infrared survey of the plane of the Milky Way (MIPSGAL) 
(Carey et al. 2009) and at 25 \micron\ from IRAS.
If these could be used to measure star formation rates in regions of
our Galaxy, it would be very useful. The goal of this paper is to test
the limits of applicability of the extragalactic relations for regions
within our Galaxy.  Since we can observe star forming 
regions in the Milky Way in more detail, testing extragalactic SFR relations on
nearby regions can also  provide some perspective on the use of such 
relations in other galaxies.  

In order to test how well 24 \micron\ emission can trace SFR, another method for
tracing SFR is needed for comparison.  We tie our measurements to 
those in nearby clouds, where we can count YSOs of a certain age. These
provide a completely independent and reasonably accurate measure of the SFR.
These nearby clouds are not forming high mass stars, which means that 
the IMF is not fully-sampled in these regions. 
Since one of the assumptions in deriving SFR from IR emission is that the IMF 
is fully-sampled in the regions, studying the use of IR tracers in these nearby clouds 
can tell us about the effect of under-sampling the IMF on SFR calibration. 
We then extend the study to regions forming massive stars.
These regions are at larger distances than the nearby clouds, and counting
individual YSOs in these regions as a measure of SFR is not applicable. 
With the lack of a direct method of measuring SFR, we instead compared SFR 
measured from 24 \micron, \lir\, and radio continuum emission.    
In section 2 we describe the
sample of star forming regions used in the study. In section 3 we describe how
the SFR was calculated for a sample of nearby molecular clouds.
In section 4, we consider high mass star forming regions using samples of
massive, dense, clumps from Wu et al. (2010).  The resulting comparison of all
the SFRs in this study is described in section 5, and we summarize the
results in section 6.

\section{The Sample}\label{sample}

Two groups of sources were included in this study. 
The first group consists of nearby molecular clouds
with evidence of low-mass star formation. This group has
the advantage of having an independent estimate of the SFR from
counting YSOs. The second group consists of massive
dense clumps with evidence of high mass star formation.
This group does not have SFRs from YSO counting, but it
is more representative of the star formation regions that might
be seen in other galaxies.

The first group consists of 20 clouds within 1 kpc of the Sun, in the
structure known as the Gould Belt (GB). They have data from Spitzer Legacy
programs and ancillary data (Evans et al. 2003, core to disk (c2d); and Allen
et al. in prep., GB). 
The clouds are listed in Table 1, along with their distances.
All the clouds have been observed in all IRAC (3.6, 4.5, 5.6, 8.0 $\mu m$ ) 
and MIPS bands (24, 70, 160 $\mu m$), using the same procedures and
data reduction methods. Young Stellar Objects
(YSOs) were identified and categorized into their SED classes (Class I, Flat, Class II, 
and Class III) using the spectral index following the criteria from Green et al. (1994).
The details on identifying YSOs and calculating
SFR in these clouds can be found in Evans et al. (2009) and Heiderman
et al. (2010). 
We also make use of data from the IRAS data archive for assessing
the large scale emission from the clouds.

The second group contains massive dense clumps with evidence of high mass
star formation, selected from Wu et al. (2010). This sample is a subsample 
of a large survey by Plume et al. (1997) of regions associated with water 
masers, which are indicators of an early phase of
massive star formation, most of which contain compact or 
ultracompact \hii\ regions. These clumps have characteristic densities from
CS excitation of about $10^{6}$ cm$^{-3}$ (Plume et al. 1997).
The mean and median virial masses are 5300 and 2700 \msun, respectively.
Most of these clumps
have been observed in many molecular line transitions, such as CS lines (Plume
et al. 1992, 1997; Shirley et al. 2003), HCN \jj10\ and \jj32\  (Wu et al. 2010), 
HCO$^+$ and several others (Reiter et al. 2011).
Some of the clumps have also been observed in 350 \micron\ dust continuum
emission by Mueller et al. (2002), who also tabulated IRAS data. 

\section{Analysis of the Regions Forming Low-mass Stars} \label{lowmass}

Emission at 
24 \micron\ has been used in many extragalactic studies as a star formation
tracer. A number of studies have derived an expression for the SFR
as a function of the 24 \micron\ emission [\sfrmir] 
(Calzetti et al. 2007, Alonso-Herrero et al. 2006, Rieke et al. 2009, 
Wu et al. 2005, Zhu et al. 2008, Relano et al. 2009, Perez-Gonzalez et al. 2006). 
Various calibrations of \sfrmir\ are compared
in Calzetti (2010).  Our goal is to 
test these relations by comparing the SFR using 24 \micron\ emission
with the SFR using YSO counting  (Evans et al.  2009, Heiderman et al. 2010). 

The YSO counting method uses the following equation.
\begin{equation}
\sfryso\ = {N({\rm YSOs}) \mean{M_*} /t_{excess}}.
\label{ysocounteq}
\end{equation}
Assuming an average stellar mass of $\mean{M_*} = 0.5 M_\odot$  
and an average time for YSOs to have an infrared excess of 
$t_{excess} = 2$ Myr, the SFRs were
calculated by Evans et al. (2009) and Heiderman et al. (2010). 
The average mass was chosen to be consistent with IMF studies
(Chabrier 2003, Kroupa 2002) 
and consistent with an average mass for some clouds although there 
may be variations between clouds (Evans et al. 2009). 
They are collected in Table 1. The largest source of uncertainty
is the lifetime of the infrared excess (perhaps $\pm 1$ Myr).

\subsection{24 \micron\ emission from YSOs}

We now compare the SFRs calculated from counting YSOs [\sfryso ] to 
the SFRs calculated using \sfrmir.  
Since 24 \micron\ emission comes from dust that has been heated by stellar
radiation and does not require high energy photons, 
it may be able to pick up the star formation rate of even low-mass YSOs.

The first step was to calculate the total 24 \micron\ emission coming
from all the YSOs in each cloud.   
The flux densities at 24 \micron\ for individual YSOs were extracted
from data bases and summed over all the YSOs in individual clouds. The
resulting total YSO flux for each cloud is shown in Table 1. 
Using the distances to the clouds (Heiderman et al. 2010, updated distances 
can be found in Dunham et al. 2012 in prep), 
the 24 \micron\ luminosity can be calculated from the total 24 \micron\ flux
density. 

From the total 24 \micron\ emission from YSOs, we computed SFR(YSO, 24 \micron).
The relation for \sfrmir\ that we used in this study came from the work of 
Calzetti et al. (2007), who adopted the starburst99 stellar synthesis model
and Kroupa's IMF (Kroupa et al. 2001) in the calibration. 
Kroupa's IMF has been used in 
many studies for calibrating SFR; it has the form and stellar mass range
 described by (Chomiuk et al. 2011, Kennicutt et al. 2009, Murphy el al. 2011):
\begin{equation}
\psi (\log(m)) \propto m^{-0.3}  (0.1 \leq m \leq 0.5 \, \msun ), \nonumber 
\end{equation}
\begin{equation}
\psi (\log(m)) \propto m^{-1.3} (0.5 \leq m \leq 100 \,  \msun).   \nonumber 
\end{equation} 
Calzetti et al. (2007) uses Kroupa's IMF but with an upper mass limit of 
120 \msun.
 The \sfrmir\ is
\begin{equation}
\text{SFR} (\msun\; \text{yr}^{-1} ) = 1.27 \times 10^{-38} [ L_{24 \, \micron}
 (\text{ergs s}^{-1}) ] ^{0.8850} ,
\label{mirequation}
\end{equation}
where $L_{24\, \mu m}$ is the total 24 \micron\ luminosity per unit
frequency  times the frequency ($\nu L_{\nu}$).
The calculated SFRs for each cloud are as shown in Table 1.

It is clear that SFR(YSO, 24 \micron) vastly underestimates \sfryso. 
The mean ratio of \sfryso\ to SFR(YSO, 24 \micron) is 1867$\pm$1335.

\subsection{Total 24 \micron\ Emission}

Since the relation in equation \ref{mirequation}
was derived for extra-galactic star formation, where individual
YSOs are not resolved, we should
expect the detected flux to be contributed from diffuse emission as well as
from point sources. 
In this section, we consider the total emission, which includes diffuse as well 
as point source emission in 
\sfrmir.

To compare SFR from the total 24 \micron\ emission with the SFR from YSO
counting, the calculations have to come from the same area of the clouds.
Boundaries for each cloud used for identifying YSOs were chosen using contours
from extinction maps. Therefore, we chose the same 
boundaries for calculating diffuse 
emission.
 All clouds' boundaries were chosen to be extinction
contours of $\av = 2$. The exceptions are Serpens and Ophiuchus 
for which the c2d survey extended down
to $\av=6$ and $\av=3$ respectively (Evans et al. 2009). 
The total flux used to calculate the SFR should also be
emission only from the clouds themselves. 
Images that cover the area inside the cloud's boundary can still contain 
foreground and background emission not associated with the clouds. To 
include only emission from the clouds, we subtracted background emission.
To do this, we needed large scale images that cover not only the area of the
cloud defined by extinction contours, but also the area surrounding the contour
boundaries. MIPS images from the Spitzer survey have good spatial resolution
but lack the area coverage needed for background estimations. Therefore, we
chose to use IRAS images for our diffuse emission analysis.

The Infrared Astronomical Satellite (IRAS) observed 96$\%$ of the sky in four
bands (12, 25, 60, 100 \micron). 
We used 25 \micron\ IRAS images from the
the Improved Reprocessing of the IRAS Survey (IRIS) obtained
from the Infrared Processing and Analysis Center (IPAC)
as a substitute for 24 \micron\ data.
First the total flux densities inside contour boundaries were 
calculated for each cloud. 
We then chose a ``sky annulus'' for each cloud separately by choosing an area
surrounding the cloud's boundary while avoiding any extended emission that
seemed to be connected to the cloud. The background level was estimated by
summing over the flux inside the sky annulus divided by the total number of pixels
to estimate the background value per pixel (Jy/pix). The total flux inside
contour boundaries minus the background flux (background flux = average 
background level per pixel $\times$ number of pixels inside the boundary)
gave the actual flux from the clouds. 
The  25 \micron\ emission coming from the clouds themselves turns out to be
very small compared to the foreground/background emission. 
The 25 \micron\ luminosities calculated from the background subtracted flux
for all the c2d and Gould's Belt clouds are shown in Table 2. 
For clouds with background emission comparable to the total emission inside the
boundaries, namely Lupus IV and Auriga North, we set the 25 \micron\ luminosities 
and \sfrmir\ to zero.  
With the 25 \micron\ luminosities, the SFR for each cloud was 
obtained using Equation \ref{mirequation}. 
The differences between luminosities measured at 24 \micron\ and 25 \micron\ 
should be quite small. 

\begin{figure}[h]
\centering
\includegraphics[scale=0.7]{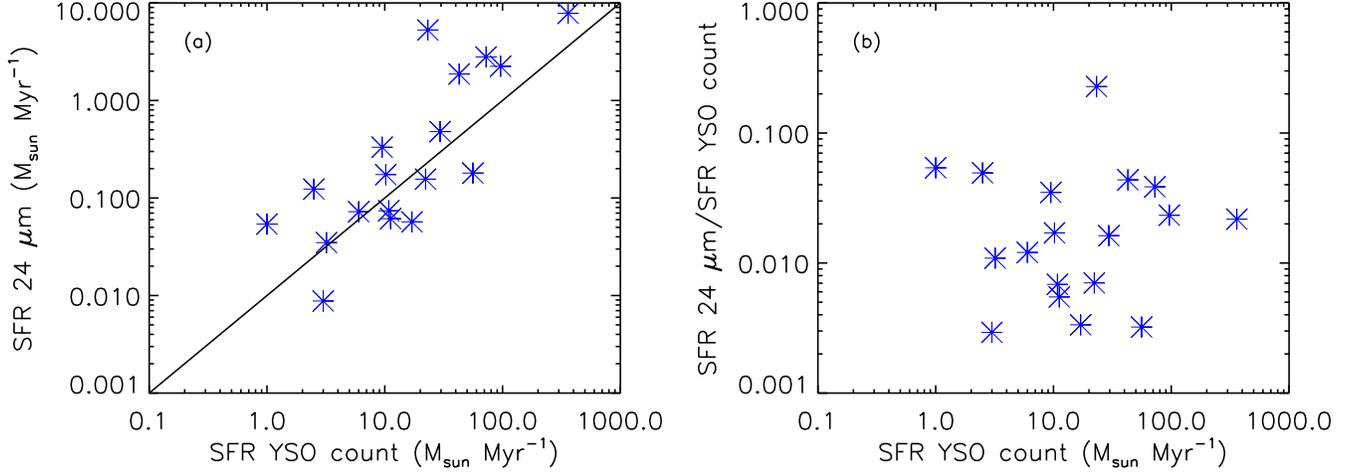}
\caption{ SFR(24 \micron) versus SFR(YSO count) for c2d and Gould's Belt clouds,
with \sfrmir\ calculated from the background-subtracted diffuse emission.
The solid black line represents a line of SFR(24 \micron) = 0.01SFR(YSO count).
\label{fig:sfr24lm}}
\end{figure}

Table 2 compares the \sfrmir, which is calculated from the total emission 
including point sources and diffuse emission, with \sfryso.
It is clear from the table that \sfrmir\ greatly underestimates
\sfryso.  The average ratio of \sfryso\ to \sfrmir\ is
$107 \pm 109$, with a median of 61.6. Figure~\ref{fig:sfr24lm}(a) shows a plot of 
\sfrmir\ over SFR(YSO count), and Figure~\ref{fig:sfr24lm}(b) shows 
a ratio of \sfrmir/\sfryso\ over \sfryso. 

\subsection{Contributions from Stellar Continuum Emission}

Calzetti et al. (2007) developed relations between SFR and 
emission at two MIR wavelengths of 8 and 24 \micron. Since only the
dust emission should measure SFR, stellar continuum emission needed to be
subtracted from the flux. The stellar continuum subtraction was performed for
the 8 \micron\ emission, but contributions to the 24 \micron\ flux from
stars was considered to be negligible. 
We used c2d clouds as sample regions to see how much stellar 
continuum contributes to the total flux. The c2d project identified all point 
sources, which include background and foreground stars, for all clouds. 
These background/foreground stars in fact dominate the source counts 
in each cloud. With the available data, we can compare the
 contributions from point sources, which can be separated into
 YSO and non-YSO, to the total 24 \micron\ flux. 
First, we calculated the flux from all identified objects in the 24 
\micron\ MIPS images. Then the flux from YSOs was subtracted from the 
all-object flux to
get the non-YSO object flux. 
In extragalactic studies, when looking at star forming regions 
the flux is the total flux emitted from the projected area. 
To see how much stellar emission contribute to total flux, 
we compare the non-YSO flux to the total flux (before background subtraction). 
 The results show that stellar continuum 
contributes very little to the total flux. The contribution is larger for some clouds, 
specifically clouds with little diffuse emission, but stellar contributions to 
the total flux are less than 10 percent for all clouds (Table 3).  

\subsection{\lir\ }

Another tracer of star formation often used in extragalactic studies is the
total infrared luminosity.  While 24 \micron\ emission arises from warm
dust grains or from small, transiently heated dust grains, most of the
emission from dust in molecular clouds peaks at a longer
wavelength, in the far-infrared.
The total infrared luminosity should then trace the bulk of the
dust emission.  With the available IRAS data, the total infrared luminosity 
(\lir) for all the c2d and GB clouds can be estimated from: 

\begin{equation}
\lir = 0.56 \times D^2 \times (13.48 \times f_{12} +5.16 \times f_{25} + 2.58
\times f_{60} + f_{100} ),
\end{equation}
where $f_i$ is the flux in each IRAS band in units of Jy, $D$ is the distance
in kpc, and \lir\ (8-1000 \micron) is in units of \lsun  (Wu et al. 2010). 
Each of the IRAS bands have a slightly different angular resolution:
3.8\arcmin, 3.8\arcmin, 4.0\arcmin, and 4.3\arcmin
 for IRIS plate of 12, 25, 60, and 100 \micron\ respectively
(Miville-Deschenes \& Lagache 2005). 
However, the angular size of our objects are in the order of a few degrees. 
We therefore did not take into account the differences in the resolutions. 
The flux in each band was computed with the same technique used for
the flux at 25 \micron, including background subtraction.

To calculate \sfrlir, we used the extragalactic relation for starburst galaxies from
K98. However, the SFR(\lir) from K98 assumed a Salpeter form of the IMF.
For consistency, all our calculations should be based on the same IMF model. 
A Salpeter IMF gives a Lyman continuum photon rate of 1.44 times higher 
than Kroupa IMF (from 0.1-100 \msun)  for the same 
SFR (Chomiuk et al. 2011, Kennicutt et al. 2009).
Assuming that \lir\ scales with Lyman continuum photon rates,
we then divided \sfrlir\ from K98 by 1.44 to obtain
\begin{equation}
\text{SFR(\msun\ year}^{-1} ) = 3.125 \times 10^{-44} L_{TIR} (\text{erg s}^{-1} ) ,
\label{lirequation}
\end{equation}
where \lir\ is the total infrared luminosity (8-1000 $\mu m$). 

\begin{figure}[h]
\centering
\includegraphics[scale=0.7]{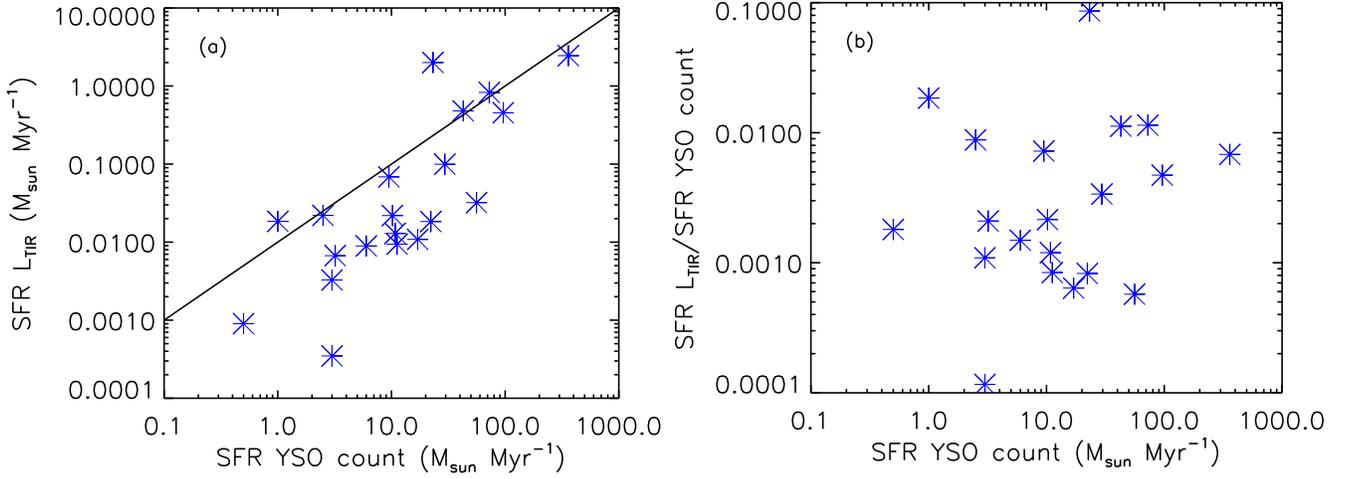}
\caption{ \sfrlir\ versus SFR(YSO count) for c2d and Gould's Belt clouds. The
solid black line represents the same line as the line in Figure 1 of \sfrlir\ =
0.01SFR(YSO count)   \label{fig:sfrlirlm}}
\end{figure}

The results (Table 2) show that \lir\  underestimates \sfryso\ 
for all the clouds, with the mean ratio of \sfryso\ to \sfrlir\
of $969 \pm 1870$ and median of 480. Figure~\ref{fig:sfrlirlm}(a) shows
\sfrlir\ over \sfryso, and Figure~\ref{fig:sfrlirlm}(b) shows 
the ratio of \sfrlir/\sfryso\ versus \sfryso. 

\begin{figure}[h]
\centering
\includegraphics[scale=0.7]{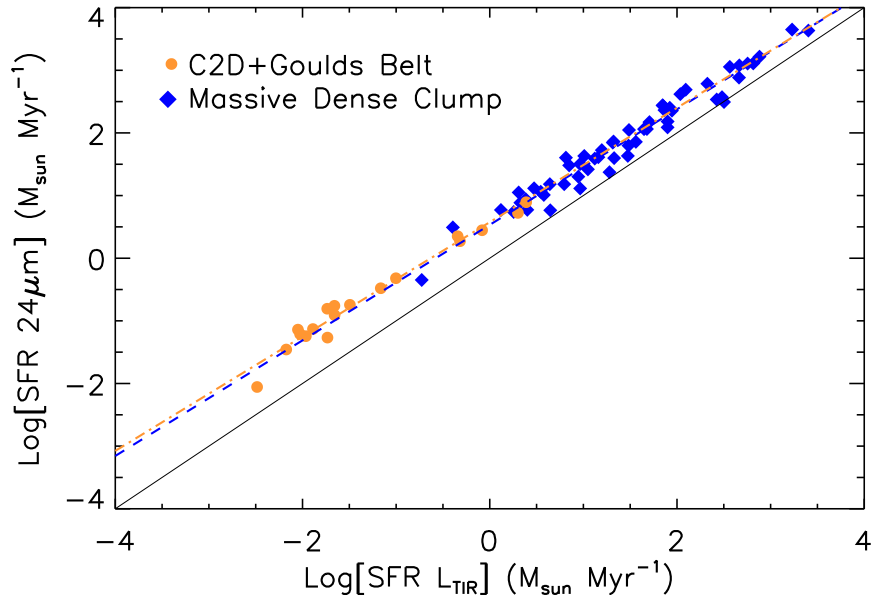}
\caption{ Log[SFR(24 \micron) versus Log[\sfrlir] for c2d, Gould's belt clouds,
and massive dense clumps. The solid black line represents a line of
\sfrmir/\sfrlir\ = 1; a dash-dot orange line represents a fit to the c2d and
Gould's Belt cloud data points; and a dot blue line represents a fit to
the massive dense clump data points.  \label{fig:sfrall}}
\end{figure}

With both the 24 \micron\ and \lir\ available, we also compared 
\sfrmir\ with \sfrlir. Figure~\ref{fig:sfrall} shows \sfrmir\ over \sfrlir\
 with the low mass star forming clouds data represented by orange circles. 
 The two SFRs agree well with each other with 
 average ratio of \sfrlir/\sfrmir\ of $0.22\pm0.08$ and a median of 
 0.33. A curved fit was performed using MPFITEXY routine 
(Williams et al. 2010; Markwardt 2009) with adopted uncertainties of 50\% 
for both SFRs. 
The solid black line represents
 a line of \sfrmir\ = \sfrlir\ while the dot-dashed, orange line 
 represents a least-square fit for the nearby clouds of 
 \begin{eqnarray}
\log[\sfrmir]  &=& (0.58\pm0.13) \\
&+& (0.91\pm0.08) \times \log[\sfrlir]. \nonumber
\nonumber
\end{eqnarray}

\section{Analysis of Regions Forming High-Mass Stars}\label{highmass}

So far we have found that the extragalactic relations between SFR
and 24 \micron\ or total infrared badly underestimate the
SFR in nearby molecular clouds, which are not forming stars of
high mass. Here we address the issue for regions forming massive stars,
using the dense clump sample discussed in \S \ref{sample}.
These clumps have an average distance
of $3.9\pm2.4$ kpc and a median of 3.5 kpc.  

\subsection{IRAS 25 \micron\ emission and total infrared luminosity \lir}

The fluxes for the IRAS bands for these clumps are available from the IRAS point
source catalog (PSC) and tabulated by Mueller et al (2002). However, most of 
the massive dense clump sources are extended sources. Examining the 
images of these sources showed that the IRAS point source catalog could 
underestimate the flux because the average source size is 
larger than the IRAS beam size. To obtain more accurate values of the flux, 
we performed photometry on the massive dense clump sample instead of 
adopting the flux from PSC. 

IRAS IRIS images in all four bands were used for photometry. 
Aperture photometry was performed on each source with the use 
of IDL routine APER and by setting  the aperture radius 
to be equal to the FWHM of a 1D gaussian fit.  
Most of the sources are in a crowded field, which complicated the 
photometry. Sky subtraction was done by choosing a sky region 
for each source by eye and averaging the flux within the region to obtain 
sky level. The result gives a flux in all four IRAS bands for a 
total of 56 sources. 

The total infrared luminosity 
and the \sfrlir\ was calculated from the same equation used in the last
section (Equation 3 and 4). 
Note that \lir\ from our photometry is higher than \lir\ from the PSC 
by a factor of 2 on average. The \sfrmir\ was also calculated 
in the same way by using the relation in Equation 2. 
Ideally, we would now compare the SFRs from infrared emission to
\sfryso as we did for low-mass regions. 
However, because of the greater distance and the presence
of diffuse emission, counting YSOs is not practical in these regions.
Without the YSO count, we cannot test the IR SFR tracers against a direct measure of SFR. 
With more than one method of tracing star formation, we can test to 
see if different tracers give consistent measures of SFRs. 

As shown in Table 4, the two IR SFRs are comparable to each other with the average
ratio of \sfrlir\ to \sfrmir\ $ = 0.41\pm0.19$. The median is 0.37.  
Figure~\ref{fig:sfrall} shows the comparison between \sfrmir\ and \sfrlir\ 
for the clumps, which is represented by blue diamonds.
The dashed, blue line represents a least-square fit for the massive 
dense clump data of
\begin{eqnarray}
\log[\sfrmir]  &=& (0.53\pm0.08) \\
&+& (0.92\pm0.05) \times \log[\sfrlir]. \nonumber
\nonumber
\end{eqnarray}

\subsection{Radio Continuum Emission}
In addition to infrared emission, radio continuum emission is also used as a 
SFR tracer for galaxies in several studies (Condon et al. 1992, Yun el al. 2001, 
Jogee et al. 2005, Murphy et al. 2011). 
For normal and starburst galaxies, most of the radio emission is 
free-free emission from ionized gas and synchrotron emission 
from relativistic electrons (Yun et al. 2001). 
Free-free emission traces ionized gas inside \hii\ regions, along
with some more diffuse emission from extended ionized gas, while 
synchrotron emission traces relativistic electrons accelerated 
by supernova remnants, which are much more widely distributed. 
Both of the sources of the radio emission are related to high mass 
star formation 
because high mass stars produce \hii\ regions 
while stars with $M\geq 8$ \msun\ 
produce core-collapse supernova (Yun et al. 2001). 
However, the quantitative relation between synchrotron emission and
star formation is less direct, being derived from a correlation
between the synchrotron and far-infrared emission 
(de Jong et al. 1985; Helou et al. 1985; Condon 1992). 

For this study, we used radio continuum as another independent source 
of SFR tracer 
for comparison with \lir\ since both radio continuum and \lir\ should 
trace the presence of high mass stars. 
In a spectrum of a whole galaxy, synchrotron emission dominates 
emission at $\nu \leq 30$ GHz (Condon et al. 1992). 
However, our samples are on much smaller scales than for extragalactic studies. 
In the absence of nearby supernova remnants, radio emission from high mass 
star forming regions is dominated by thermal free-free emission. 
To use radio continuum as a SFR tracer for the massive dense clump samples 
we need to connect free-free emission to a total number of massive stars.  
Thermal (free-free) luminosity is related to the rate of photoionizing photons 
(Lyman continuum photons) by
\begin{eqnarray}
\left( \frac{N_{UV}}{\text{phot s}^{-1}} \right) \geq 6.3 &\times& 10^{52} \left( \frac{T_e}{10^4 \text{K}} \right)^{-0.45} 
\left(  \frac{\nu}{\text{GHz}} \right)^{0.1} \\ \nonumber
 &\times& \left( \frac{L_T}{10^{20}\text{W Hz}^{-1}} \right), 
\end{eqnarray}
where $N_{UV}$ is the production rate of Lyman continuum photons per second, 
$T_e$ is the electron temperature, $\nu$ is the frequency, and $L_T$ is 
the thermal emission luminosity, assuming it is optically thin in this part of 
the spectrum (Condon et al. 1992). 
Using Kroupa's IMF and stellar spectral model from Starburst99 (Leitherer et al. 1999), 
the rate of photoionizing photons is related to SFR by (Chomiuk et al. 2011)
\begin{equation}
\frac{\text{SFR}}{\msun \; \text{yr}^{-1}} = 7.5 \times 10^{-54} \left( \frac{N_{UV}}{\text{phot s}^{-1}} \right) . 
\end{equation} 
We get 
\begin{equation}
\frac{\text{SFR}}{\msun \; \text{yr}^{-1}} = 0.47 \left( \frac{T_e}{10^4 \text{K}} \right)^{-0.45} \left( \frac{\nu}{\text{GHz}} \right)^{0.1} 
\left( \frac{L_T}{10^{20} \text{W Hz}^{-1}} \right).  \nonumber
\end{equation} 
For an electron temperature of T$_e \sim 10^4$ K, 
the thermal radio SFR relation is
\begin{equation}
\frac{\text{SFR}}{\msun \; \text{yr}^{-1}} = 0.47 \times 10^{-20} \left( \frac{\nu}{\text{GHz}} \right)^{0.1} \left( \frac{L_T}{\text{W Hz}^{-1}} \right). 
\end{equation}

For the radio continuum data, we used radio surveys that cover 
the regions of the Galactic plane that coincide with the massive dense 
clump sample. The radio data in this study was obtained from two surveys. 
The first set of data came from a survey of the Galactic plane at 4.875 GHz 
by Altenhoff et al. (1979; hereafter A79). The radio data were obtained with 
the 100-m Effelsberg with a half-power beamwidth of 2.6$\arcmin$ over 
the galactic longitude range of $l=357.5^\circ$ to $60^\circ$ and 
galactic latitude of $b=\pm 2^\circ$. The second set of radio data were 
obtained from an earlier survey by Altenhoff et al. (1970; hereafter A70). 
The survey of the Galactic plane at 1.414, 2.695, and 5.000 GHz 
covered a range of $l= 335^\circ$ to $75^\circ$ 
and $b=\pm 4^\circ$ with a half-power beamwidth of 
approximately $11\arcmin$. The observations for the three wavelength 
bands were made with the 300-ft transit paraboloid antenna at the NRAO, 
the 140-ft antenna at NRAO, and the 85-ft parabolic antenna at Fort Davis 
for 1.414, 2.695, and 5.000 GHz respectively (Altenhoff et al. 1970). 
Using the 4.875 GHz (A79) survey has the advantage of having a comparable 
resolution to the infrared data from IRAS ($2.6\arcmin$ for A79 and 
$\sim 2\arcmin$ for IRAS 100 \micron), making it suitable for 
comparison between radio and infrared data. 

We first matched objects from the radio surveys to the massive dense clump 
objects by matching their coordinates. The matching objects have center 
coordinates within a few arcminutes of each other. Lockman (1989) provides 
radio recombination line data for these radio sources from his survey of radio 
\hii\ regions in the northern sky. We compared radio recombination line 
velocities of matched objects to line velocities (HCN \jj10, \jj32\ and 
CS \jj21, \jj76) from Wu et al. (2010). 
We kept the objects with velocities approximately 
within $\pm5$ \kms\ between the two data sets. Our matching resulted in a 
total of 22 objects with available radio continuum flux, radio recombination line
velocity, and infrared luminosity. 

A79 provides a peak intensity for each radio source along with a FHWM. 
The integrated flux for each object was calculated for a total of 18 objects 
by assuming a Gaussian profile for both the source flux distribution and the 
beam profile. A70 provides integrated flux and FWHM data for an additional 4 
objects. 
Then SFR(radio) was calculated from Equation 7. After obtaining SFR(radio),
our next step was to compare them to IR SFR. However in order to compare 
radio data to infrared data, the two sets of data should come from 
equal areas of the objects. Aperture photometry was performed 
on IRAS IRIS images with a chosen aperture radius equal
 to the radio FWHM size of each object. The aperture size was chosen 
 to capture most of the infrared flux of the objects 
 without contamination from other nearby sources 
and to make the observed areas comparable to those 
of the radio data. 

\begin{figure}[h]
\centering
\includegraphics[scale=0.7]{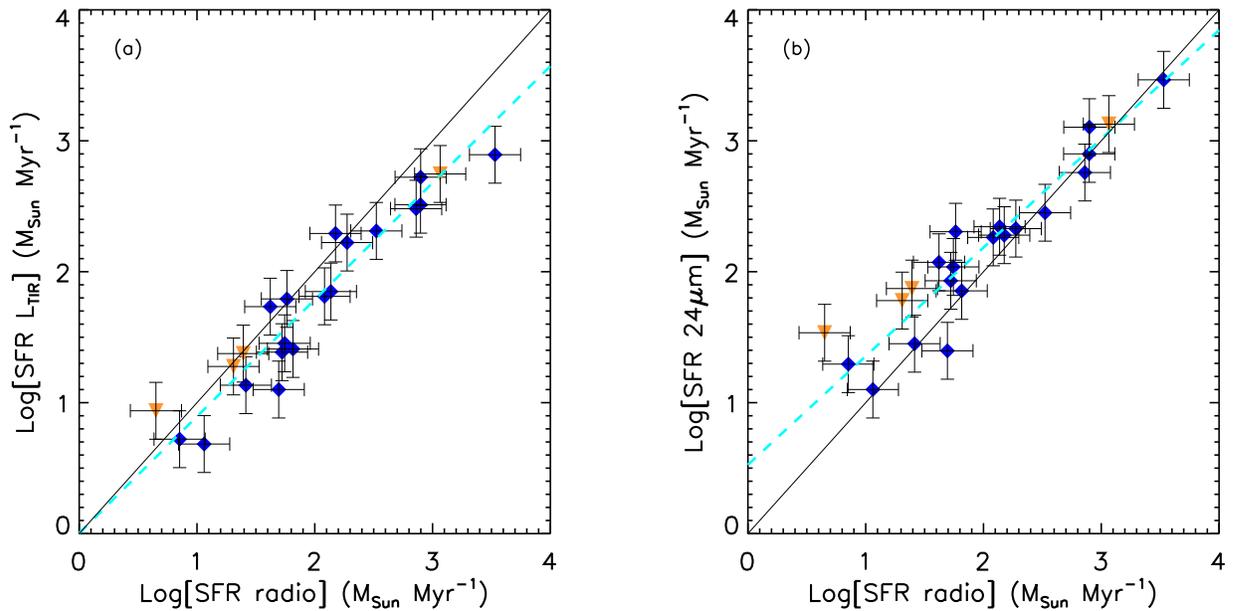}
\caption{ \sfrlir\ versus SFR(radio) for massive dense clumps. The blue squares represent 
data from A79, and the orange triangles represent data from A70. The solid, black line represents 
a line where the two SFRs are equal while the blue, dashed line represents a fit of 
log[\sfrlir] = 0.0029+0.89 Log[SFR(radio)].  \label{fig:sfrradio}}
\end{figure}

The resulting SFR(radio), \lir, \sfrlir\, and \sfrmir\ are included in Table 5.  
SFR(radio) and \sfrlir\ are well correlated, with an average 
ratio of SFR(radio)/\sfrlir\ of $1.8\pm0.8$, a median of 1.9, 
and a linear correlation coefficient of 0.90. 
There are many sources of uncertainties in our calculations of SFR, which 
makes it difficult to estimate realistic errors for each source. 
We instead adopted a 50$\%$ uncertainties for both SFRs 
and performed a curve fit using MPFITEXY routine 
(Williams et al. 2010; Markwardt 2009). 
Figure~\ref{fig:sfrradio}(a) shows \sfrlir\ versus SFR(radio) 
with a solid line representing SFR ratio of one and a 
dashed line representing a best fit to the data of 
\begin{eqnarray}
\log[\text{\sfrlir}] &=& (0.0029\pm0.18) \\
&+& (0.89\pm0.085) \times\log[\text{SFR(radio)}]. \nonumber
\end{eqnarray}

SFR(radio) and \sfrmir\ are also well correlated with an 
average ratio of SFR(radio)/\sfrmir\ of $0.76\pm0.42$, a 
median of 0.79, and a linear correlation coefficient of 0.98.
Figure~\ref{fig:sfrradio}(b) shows \sfrmir\ versus SFR(radio) 
with a dashed line representing a best fit of
\begin{eqnarray}
\log[\text{\sfrmir}] &=& (0.53\pm0.17) \\
&+& (0.83\pm0.08) \times\log[\text{SFR(radio)}]. \nonumber
\end{eqnarray}
\\

\section{Discussion}
\subsection{Low mass SF}
From the results for c2d and Gould's Belt survey, it is clear that the
SFRs from 24 \micron\ do not agree well with SFRs from YSO counting. 
First of all,
24 \micron\ emission from YSO point sources contributes very little to the total
emission of the clouds. Even when we included the diffuse emission into
our calculation of \sfrmir, the resulting values are still much lower 
(by a factor of about 100 than \sfryso). Nonetheless, we can ask whether
there is any relation at all between \sfrmir\ and \sfryso.
Figure~\ref{fig:sfr24lm}(a) shows a plot of \sfrmir\ versus \sfryso. 
The solid black line represents  a ratio of 100.
The figure shows that there is a general correlation between the two 
with the Pearson linear correlation coefficient of 0.83.
Perhaps the 24 \micron\ emission might provide a rough guide
to the SFR, but with a different conversion factor.
However, the scatter is large.
Figure~\ref{fig:sfr24lm}(b) shows the ratio of SFR(24,diffuse)/SFR(YSO count). 
The discrepancies and scatter between the two SFRs 
persists throughout the range of SFRs.
A similar result was obtained for the comparison of \sfrlir\ with 
\sfryso, as shown in Figure ~\ref{fig:sfrlirlm}.  There is again a weak correlation with 
a correlation coefficient of 0.77, but the underestimate of \sfryso\ is even greater.
The solid black line represents the same line of $\sfryso = 100 \times \sfrlir$,
as shown in Figure~\ref{fig:sfrlirlm}(a). 

The disagreement between SFR(IR) and \sfryso\ is not surprising since 
these clouds are not forming very massive stars, which would dominate the 
luminosity if the IMF is fully sampled. The undersampling of the IMF along with 
other possible causes behind the discrepancy in SFRs are discussed below. 

\subsubsection{External Heating}
\begin{figure}[h]
\centering
\includegraphics[scale=0.5]{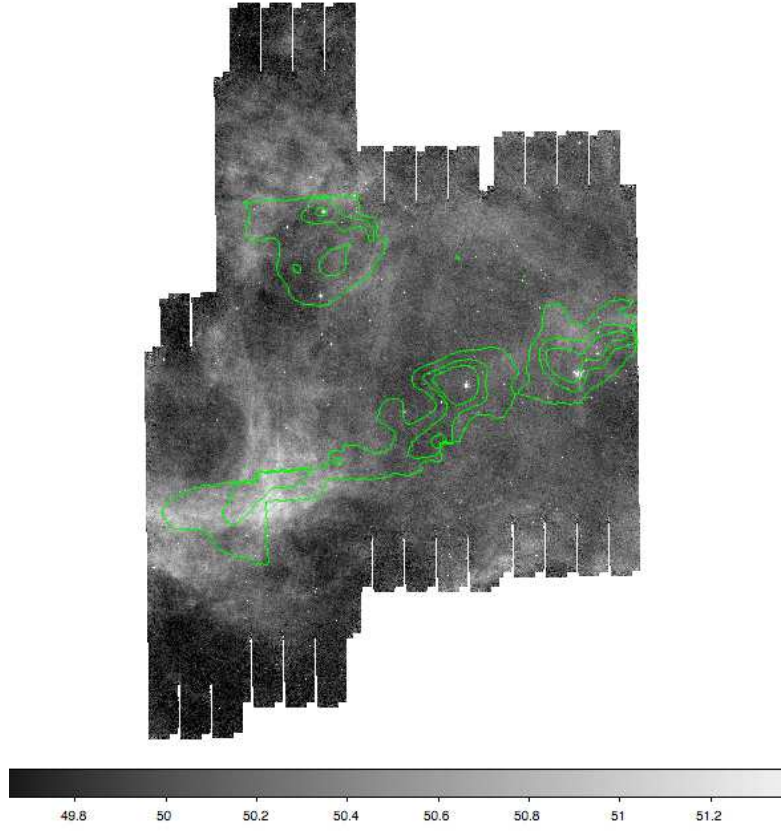}
\caption{MIPS 24 \micron\ image of Lupus I cloud with contours of
 $\av  = 2$, 4, and 6 mag in green.
 \label{fig:lupus}}
\end{figure}

\begin{figure}[h]
\centering
\includegraphics[scale=0.4]{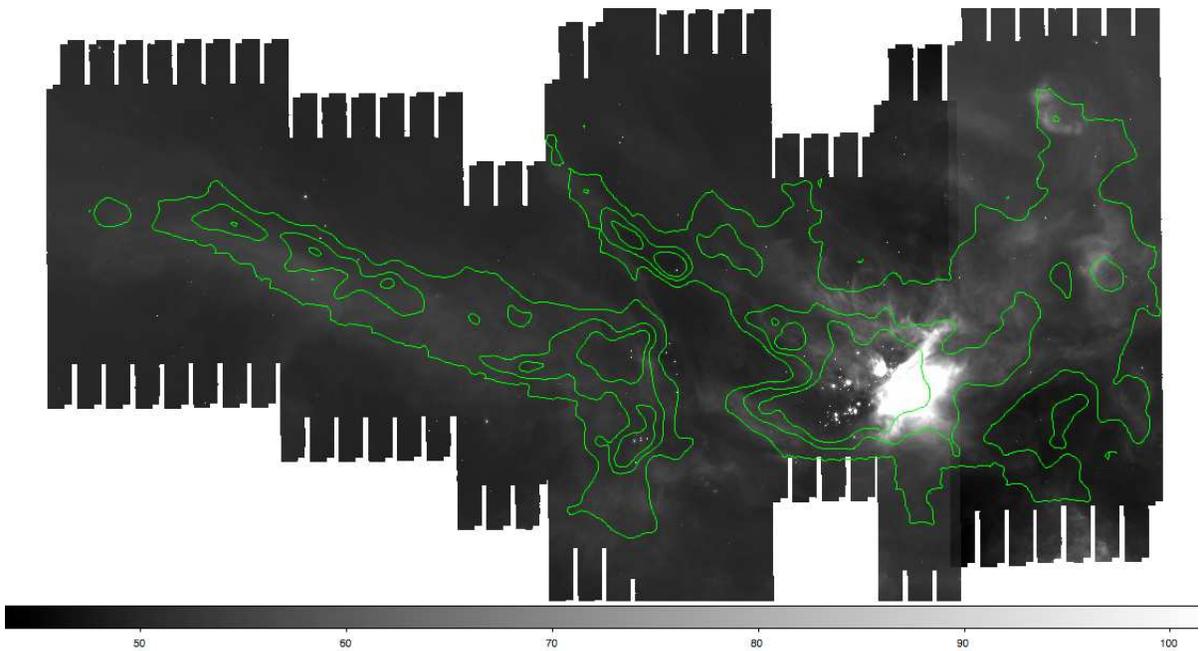}
\caption{ MIPS 24 \micron\ image of Ophiuchus cloud with contours of 
$\av  = 2$, 6, and 10 mag in green.
  \label{fig:oph}}
\end{figure}

As discussed earlier, the total fluxes from the actual clouds 
are generally small fractions of the total emission toward the regions, 
which means that a lot of the emission is background emission. 
Furthermore, much of the diffuse emission that is associated with the cloud
does not correspond to regions of high extinction or intense ongoing star
formation. As examples, Figure ~\ref{fig:lupus} and Figure ~\ref{fig:oph} 
show the images for Lupus I and Ophiuchus, with 
extinction contour levels overlaid. 
In Lupus I, the diffuse emission at 24 \micron\ is located away
from the regions of current star formation.
In contrast, in Ophiuchus, most of the diffuse emission is associated
with the cluster of forming stars spatially, and the excitation peaks
on embedded early-type stars (Padgett et al. 2008, see, Fig. 2).
In the case of the Perseus cloud, much of the diffuse 24 \micron\
emission comes from regions heated by a star lying behind the cloud 
(unrelated to current star formation) or from the IC348 cluster
(related to recent star formation) (Rebull et al. 2007).
Such differences from cloud to cloud will introduce large scatter into
the relations. In the absence of high mass stars in these clouds, 
external sources of heating could dominate the infrared emission. 

The IRAS 100 \micron\ images show more correlation with the extinction contours
than the 25 \micron\ images. 
The contribution to the \lir\ is also larger from the
100 \micron, which is closer to the peak of the general dust emission 
from molecular clouds. The resulting \lir\ may then trace the
amount of dust inside the clouds as opposed to star formation in the clouds. 
Then the correlation in Figure ~\ref{fig:sfrlirlm} could be a secondary effect of the
correlation of SFR with amount of dust for the cloud as a whole.

\subsubsection{Undersampled IMF}
Since these clouds are not forming very massive stars, clearly there
are no stars to populate the high-end of the IMF. The lack of high-mass 
stars means that it requires more mass in the form of lower-mass stars
to produce a certain luminosity than if the IMF is fully-sampled. Using SFR relations 
derived by assuming the full IMF will then underestimate the SFR in these regions. 

To see how much this affects the discrepancies in the SFRs, we looked at the
details of the SFR calibrations. Calzetti et al. (2007) calibrated the SFR-24 \micron\ 
relation by empirically fitting L(24 \micron) to \halpha. \halpha\ was then connected to SFR 
through a stellar population model assuming Kroupa's IMF, solar metallicity, and a 
constant SFR over a timescale of 100 Myr. Any differences in the IMF would have an effect
on the two steps: SFR-\halpha\ (or directly related, N$_{UV}$) relation and 
\halpha\ - 24 \micron\ ratio. We performed a test by running starburst99
with the same IMF but with a different upper limit on the stellar mass (M$_{upper}$).
We also assume that a constant fraction of the bolometric luminosity (L$_{bol}$) is being re-emitted 
in the 24 \micron\ band. 

Taking the Perseus molecular cloud as an example, the highest mass star in the cloud 
is an early B star (Rebull et al. 2007). We set M$_{upper} =15$ \msun\ and a
 constant SFR over 100 Myr. The results showed an
 underestimation of \sfrmir\ by a factor of 2.1 when assuming a full IMF.  
For \sfrlir, the relation in equation \ref{lirequation} was derived from assuming 
that all of L$_{bol}$ is re-emitted in the infrared so that L$_{bol}$=\lir. 
L$_{bol}$ was connected to SFR directly from the stellar synthesis model. 
This would result in the same underestimation of \sfrlir\ by a factor of 2.1. 

A factor of 2 difference from the cut-off IMF is still much less 
than the observed discrepancies in \sfryso/\sfrmir\ of a factor of 43 and \sfryso/\sfrlir\ 
of 210 in Perseus. 
The effect of under-sampling the IMF on underestimating the SFR will be greater 
for clouds with lower M$_{upper}$. For many clouds M$_{upper}$ is even lower than 
15 \msun. We tested the model with M$_{upper}=5$ \msun, which showed an 
underestimation of SFR by a factor of 10. 
Even with the lower M$_{upper}$, undersampled IMF still cannot account for the large 
discrepancies in the whole sample. 
We tested the the effect of under sampling IMF by changing M$_{upper}$, but
in regions of low SFR stochastic sampling of the IMF could also be important, 
especially in contributing to the scatter in the sample (da Silva et al. 2011, 
Eldridge 2012). 

\subsubsection{Star Formation Timescale}
The time scale of constant star formation assumed in the SFR relations
is 100 Myr, much longer than a lifetime of an average molecular cloud
(few $\times 10^7$ Yr; McKee \& Ostriker 2007, Murray 2011) 
or the time scale over which YSO counting is relevant ($\approx$ 5 Myr).
On a longer time scale the contribution of high mass stars to the
total luminosity will get smaller since low mass stars will outlast the
short-lived high mass stars. On the time scale of molecular clouds,
not accounting for the lack of massive stars will lead to even greater
underestimations of SFR than on a longer time scale.
Taking an average age of the clouds to be 10 Myr, the model results from
combining the cut-off IMF (M$_{upper}$=15 \msun) and
the change in time scale showed a higher SFR by a factor of 9.9, still lower
than the observed differences in Perseus. 
Combining the change in time scale to 10Myr and a cut-off IMF of 
M$_{upper}$=5 \msun gave a higher SFR by a factor of ~ 110, close 
to the average discrepancy in our data. 

Additionally, the assumption that all of the bolometric luminosity 
is being re-emitted in the infrared might not be valid in these regions. 
If the fraction of energy emitted in the infrared or 24 \micron\  band over 
L$_{bol}$ is not constant or is lower in regions with low SFR than in 
the regions used in the SFR calibration, then this would be another cause for
underestimation of the SFR.  

\subsection{High mass SF}
\subsubsection{\lir\ and 24 \micron}
Limited resolution, extinction, and the confusing effects of diffuse
emission prevent accurate star counts for the massive dense clumps.
Instead, we calculated the SFR 
from both 25 \micron\ and total infrared emission. 
There is a good correlation between \sfrmir\ and \sfrlir. 
Ideally, this would mean that both 24 \micron\ and \lir\ can trace
SFR well in high mass star forming regions.
However without an absolute SFR for comparison, we cannot tell
if the SFR from both tracers are accurate or if the calibration is off 
by some factor. 
Moreover, the correlation could also result if all the
clumps have similar SEDs. 

One way to distinguish these explanations is to
compare \sfrmir\  and \sfrlir\ in low mass star forming clouds.
If they show a strong correlation even when both fail to represent 
accurate SFR, the explanation of similar SEDs is likely. 
\sfrmir\ is plotted versus \sfrlir\ for both the massive dense clump 
sample and the nearby cloud sample in Figure~\ref{fig:sfrall}. 
The solid black line
represents a line of \sfrmir/\sfrlir\ = 1. 
For both data sets, \sfrmir\ is higher than \sfrlir\ on average 
with the average ratio of \sfrmir/\sfrlir\ higher for the nearby
cloud sample than for high mass sample.
The dashed red line represents a fit for the nearby clouds
while the dash-dot, green line represent a fit for the massive dense clump data.
The fact that both fit similar relationships, even though we know that
neither \sfrlir\ nor \sfrmir\ is accurately tracing SFR in the nearby
clouds suggests that the correlation is mostly driven by the similarity
of the SEDs.

The nearby cloud sample shows a smaller scatter in the data than the
high mass sample. The smaller scatter in the low mass sample suggests that
the SED for low mass star forming clouds are more uniform that those
of  massive dense clumps. If the diffuse dust continuum emission is
dominated by grains responding to the generally interstellar radiation 
field, the SED would be fairly uniform. In regions forming massive
stars, the dust energetics could instead be dominated by luminous
sources internal to the cloud, and the SED would depend more on the
distribution of luminosities of the sources and the geometry.

\subsubsection{IR and radio continuum}
After comparing \sfrmir\ to \sfrlir\ , we then compared them to SFR(radio). 
The thermal radio emission comes from a different mechanism than the infrared
emission. While infrared emission mostly traces dust surrounding \hii\ regions, 
thermal radio traces ionized gas inside \hii\ regions. 
Radio data then provides a more independent tracer of SFR
in a different part of the spectrum. The result shows that SFR(radio) also 
correlates very well with \lir\ with a correlation coefficient of 0.90. 
Radio data gives a slightly larger SFR than does \lir, as shown in Figure
~\ref{fig:sfrradio}(a), where a solid line represents a SFR ratio of one and the 
dashed line represents a best fit. 
Similarly, 24 \micron\ also correlates well with radio data as shown in Figure
!\ref{fig:sfrradio}(b). In the are of 24 \micron\, the SFR(radio) is slightly lower than 
\sfrmir\ on average. 
The fact that SFR(radio) and \sfrlir\ are comparable to each other could
indicate that both radio and infrared emission originate from the same 
source of heating, namely photons from high mass stars. 

The radio and infrared data also imply a good correlation between 
\lir\ and radio luminosity. As seen from many previous studies, 
FIR-radio correlation have been well observed among galaxies 
with a wide luminosity range and spatial scales 
(Murphy et al. 2006, Dumas et al. 2011, Hughes et al. 2006, 
Tabatabaei et al. 2007, Zhang et al. 2010). It is interesting that
even though radio continuum emission in galaxies is
dominated by synchrotron emission instead of free-free emission, 
our results still show  
that the correlation between TIR and radio emission 
extends down to parsec scales in high mass star forming regions. 

\subsection{Combining Both Samples}
\begin{figure}[h]
\centering
\includegraphics[scale=0.7]{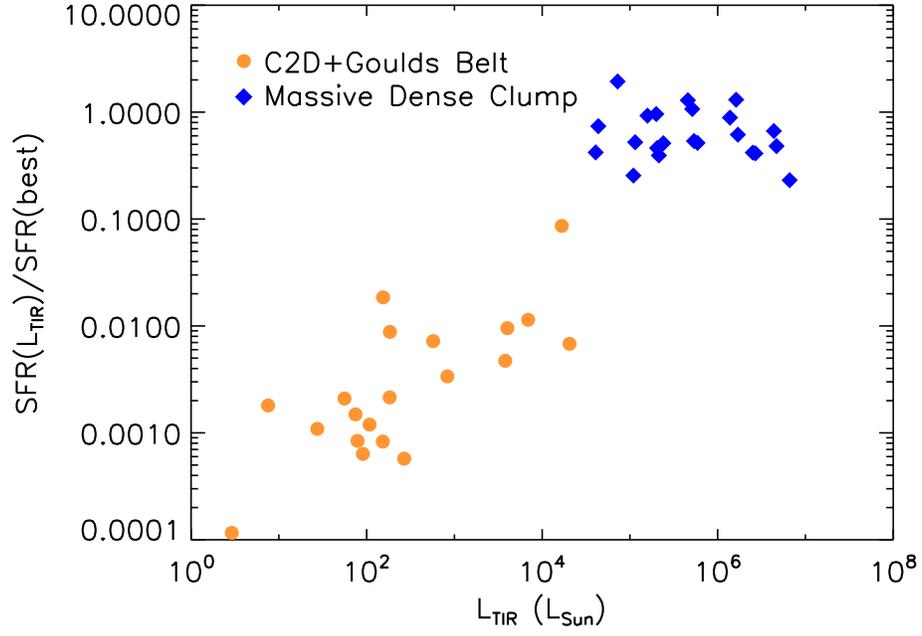}
\caption{ \sfrlir/SFR(best) versus \lir\, where SFR(best) refers to SFR(YSO count) 
for low mass regions and SFR(radio) for high mass regions. Blue stars represent 
low mass clouds (c2d+GB) and orange stars represent high mass regions 
(massive dense clump).   \label{fig:bsfr}}
\end{figure}

Our results indicate that \lir\ underestimates SFR by a large factor for low mass
regions while \lir\ gives consistent (within a factor of 2) 
SFR with SFR(radio) for high mass regions. 
Figure~\ref{fig:bsfr} shows the ratio of \sfrlir/SFR(best) 
for both low mass and high mass regions. 
SFR(best) refers to SFR(YSO count) for low mass regions and SFR(radio)
for high mass regions. 
We note that \sfryso\ is a more direct measurement of current SFR 
than SFR(radio), which depends on certain assumptions that 
went into the calibration. With the lack of \sfryso\ for high mass regions, 
we use SFR(radio) as a comparison.
The blue stars, which represents low mass clouds, show 
a general trend between the SFR ratio and \lir. \sfrlir\ is closer to the 
SFR(YSO count) at higher \lir. 
\lir\ traces SFR better for \lir\ closer to $\approx 10^{4.5}$ \lsun, which is a
transition between regions forming low-mass and regions forming 
high mass stars. 
If SFR(radio) gives an accurate measure of SFR, then the results
would mean that \sfrlir\ is a good tracer above $10^{4.5}$ \lsun.
This result would be consistent with the suggestion by Wu et al. (2005b)
that the \lir\ traces star formation above that luminosity.
Resolving YSOs in regions forming high mass stars is a next important 
step in further understanding of the use of these tracers. 

The failure of \sfrmir\ and \sfrlir\ to accurately trace SFR in nearly all the
nearby clouds has some interesting implications. 
An observer in another galaxy using \halpha\ or radio continuum emission  
would miss all star formation in a 300 pc radius of the Sun;  we
find that using 24 \micron\ emission would underestimate the 
local star formation by a factor of about 100. If the local volume
were representative of most star formation in galaxies, the SFRs
would be vastly underestimated. The fact that the same extragalactic
observers would get the {\it global} SFR in the Milky Way right
to a factor of about 2 (Chomiuk and Povich 2011) 
indicates that most star formation in
the Milky Way occurs in regions forming massive stars, 
but this might not be the case in the outer parts of the galaxies.

Finally, we note that the apparently good correlation of two
purported tracers of star formation, even in regimes where neither
is accurate, serves as a warning about accepting ``consistency'' as
evidence of accuracy.

\section{Summary}

We studied two groups of star forming clouds in the Milky Way: 20 nearby 
clouds from Spitzer c2d and Gould Belt Legacy surveys;
and 32 massive dense clumps that are forming massive stars. 
We determined the total diffuse 24 \micron\ emission for each cloud
and calculated the corresponding SFR using the relation from Calzetti et al. (2007). 
Comparing 24 \micron\
images with extinction maps shows that a significant portion of 24 \micron\
emission does not come from star-forming regions in some clouds.
We calculated the total infrared emission from the IRAS data and the
corresponding SFR. For massive dense clumps, 
we also obtained radio continuum data and calculated SFR(radio) for a total 
of 22 clumps. Then the resulting SFRs were compared with SFRs calculated using the 
method of counting number of YSOs for the nearby clouds. 
We compared \sfrlir\ with \sfrmir\ and SFR(radio) for massive dense clumps. 
The comparison shows quite a good correlation between the three SFR tracers for the
massive dense clumps, which are high-mass star forming regions, with the
average ratio of \sfrlir/\sfrmir\ = 0.6$\pm0.6$ and 
SFR(radio)/\sfrlir\ = 1.8$\pm0.9$.

Neither \sfrmir\ nor \sfrlir\ trace the \sfryso\ accurately in the
nearby clouds, where we can calibrate with an independent method.
There is a weak correlation between both tracers and \sfryso, but
a very different calibration value would be needed, and the scatter
is large.  Both 24 \micron\ and \lir\ severely
underestimate SFR for the nearby clouds. \sfrlir\ shows better agreement 
to \sfryso\ for clouds with higher luminosity. 

We would like to thank G. Helou for suggesting this study 
and the referee for suggestions that improved the work.
We would also like to thank Mike Dunham and Amanda Heiderman
for helpful discussions. we acknowledge support from
NSF Grant AST-1109116 to the University of Texas at Austin.

\clearpage

\begin{deluxetable*}{lrrrrrrrrrrrr}
\tabletypesize{\scriptsize}
\tablecaption{SFRs for the c2d and Gould's Belt clouds \label{tbl-1}}
\tablewidth{0pt}
\tablehead{
 \colhead{Cloud} & \colhead{Distance} & \colhead{N{YSOs}} & \colhead{YSO 24 \micron\ Flux}
& \colhead{SFR (YSO, 24 $\mu m$)}  & \colhead{SFR (YSO count)} &
\colhead{SFR(YSO count)/SFR(YSO, 24 $\mu m$)} \\
 \colhead{-} & \colhead{(pc)} &{-} & \colhead{(Jy)}  &  \colhead{($M_\odot$
Myr$^{-1}$ )} & \colhead{($M_\odot$ Myr$^{-1}$ )}  & \colhead{-} \\
 }
\startdata
Cha II & 178 & 24 & 7.93 & 0.0066 & 6.0 & 910 \\
Lup$^1$ & 150 & 93 & 9.45 & 0.0057 & 23 & 4000 \\
Oph & 125 & 290 & 94.2 & 0.031 & 73 & 2400 \\
Per & 250 & 385 & 77.1 & 0.090 & 96 & 1100 \\                                                                                                                         
Ser & 260 & 224 & 56.7 & 0.073 & 57 & 770 \\
Aur$^2$  & 300 & 173 & 26.5  & 0.048  & 43 & 900 \\
Cep & 300 & 118 & 24.5 & 0.045  & 30 & 670 \\
Cha III	& 200 & 4 & 0.254  & 0.00038  & 1.0 & 2600 \\
Cha I  & 200 & 89 & ...   & ... & 22 & ... \\
CrA 	& 130 & 41 & 11.9  & 0.0054  & 10 & 1900 \\
IC5146$^3$ & 950 & 131 & 16.9  & 0.25  & 33 & 130 \\
Lup VI & 150 & 45 & 6.67  & 0.0042 & 11 & 2600 \\
Lup V & 150 & 43 & 5.14  & 0.0033  & 11 & 3300 \\
Mus & 160  & 12 & 0.839  & 0.00075 & 3.0  & 4000 \\
Sco & 130 & 10 & 8.88   & 0.0042  & 2.5 & 600 \\
\enddata

\tablenotetext{1}{Combined Lup I, Lup III, and Lup IV}
\tablenotetext{2}{Combined Aur and Aur N}
\tablenotetext{3}{Combined IC5146E and IC5146NW}
\end{deluxetable*}

\begin{deluxetable*}{lrrrrrrrcrl}
\tabletypesize{\scriptsize}
\tablecaption{SFRs from diffuse emission of c2d and Gould's Belt surveys
\label{tbl-2}}
\tablewidth{0pt}
\tablehead{
 \colhead{Cloud} & \colhead{Dis} & \colhead{L$_{25\mu m} ^1$} & \colhead{\lir$^1$}  &
\colhead{SFR (YSO count)} & \colhead{\sfrmir} & \colhead{\sfrlir}
 & \colhead{SFR(YSO)/\sfrmir} & \colhead{SFR(YSO)/\sfrlir} \\
 \colhead{-}  & \colhead{(pc)} & \colhead{(L$_{\odot}$)}  &  \colhead{(L$_{\odot}$)} &
\colhead{(\msun\ Myr$^{-1}$)}  & \colhead{($M_\odot$ Myr$^{-1}$)} &
\colhead{(\msun\ Myr$^{-1}$)} & \colhead{-} & \colhead{-} \\
 }
\startdata
Cha II & 178 &14.8 & 74.7 & 6.0 & 0.0724 & 0.00893 & 83 & 670 \\
Lup I  & 150 & 6.47 & 56.0 & 3.2 & 0.0349 & 0.00670 & 92 & 480\\
Lup III  & 200 & 11.3 & 90.7 & 17.0 & 0.0569 & 0.0108 & 300 & 1600\\
Lup IV & 150 & 0.00 & 2.91 & 3.0 & 0.00 & 0.000348& - & 8600\\
Oph	 & 125 & 917 & 6925 & 72.5 & 2.79 & 0.828 & 26 & 88\\
Per 	 & 250 & 715 & 3796 & 96.2 & 2.24 & 0.454 & 43 & 210 \\
Ser	 & 260 & 41.4 & 268 & 56.0 & 0.180 & 0.0321 & 310 & 1700 \\
Aur N &  300 & 0.00 & 7.54 & 0.5 & 0.00 & 0.000902 & - & 550 \\
Aur  & 300 & 582 & 4017 & 42.8 & 1.87 & 0.480 & 23 & 89 \\
Cep	 & 300 & 125 & 832 & 29.5 & 0.479 & 0.0995 & 62 & 300 \\
Cha III & 200 & 10.6 & 154 & 1.0 & 0.0540 & 0.0185 & 19 & 54 \\
Cha I & 200 & 35.2 & 153 & 22.2 & 0.156 & 0.0184 & 140 & 1200 \\
CrA	 & 130 & 39.8 & 183 & 10.2 & 0.174 & 0.0219 & 59 & 470 \\
IC5146E & 950 & 1882 & 16725 & 23.2 & 5.28 & 2.00 & 4.4 & 12 \\
IC5146NW & 950 & 82.5 & 573 & 9.5 & 0.332 & 0.0685 & 29 & 140 \\
Lup VI & 150 & 12.3 & 78.8 & 11.2 & 0.0614 & 0.00942 & 180 & 1200 \\
Lup V & 150 & 15.1 & 108 & 10.8 & 0.0740 & 0.0129 & 146 & 840 \\
Mus	 & 160 & 1.36  & 27.4 & 3.0 & 0.00879 & 0.00327 & 340 & 920 \\
Sco & 130 & 27.0 & 184 & 2.5 & 0.123 & 0.0220 & 20 & 110 \\
Ser-Aqu & 260 & 2938 & 20493 & 360.0 & 7.83 & 2.45 & 46 & 150 \\
\enddata

\tablenotetext{1}{ These are luminosities inside extinction contours of $\av = 2$ (
$\av = 6$ for Serpens and $\av = 3$ for Ophiuchus) after background
subtraction.}
\end{deluxetable*}


\begin{table}
\begin{center}
\caption{Comparison of different sources of 24 \micron\ emission.
\label{tbl-3}}
\begin{tabular}{crrrr}
\tableline
\tableline
 Cloud & Total Flux & Total flux after background subtraction & YSOs Flux & Non-YSO Objects Flux \\
 - & (Jy) & (Jy) & (Jy) & (Jy) \\
\tableline
Per  &  5000 & 2930 & 77.1 & 81.7 \\
Cha II & 508 & 119 & 7.93 & 17.9 \\
Oph & 18100 & 15000 & 94.2 & 223 \\
Ser & 775 & 157 & 56.7 & 47.5 \\
Lup & 1406 & 146 & 9.45 & 110 \\
\tableline
\end{tabular}
\end{center}
\end{table}

\begin{deluxetable*}{lrrrrrrrrrrr}
\tabletypesize{\scriptsize}
\tablecaption{Massive Dense Clump Sample \label{tbl-4}}
\tablewidth{0pt}
\tablehead{
 \colhead{Object} & \colhead{Distance} & \colhead{L$_{25\mu m}$} & \colhead{
Log(\lir)} & \colhead{\sfrmir} &  \colhead{\sfrlir} &
\colhead{\sfrlir/\sfrmir} \\
 \colhead{-} & \colhead{(kpc)} & \colhead{(L$_\odot$)} & \colhead{(L$_\odot$)}  &
\colhead{(\msun\ Myr$^{-1}$ )} & \colhead{(\msun\ Myr$^{-1}$ )} & \colhead{-}  \\
 }
\startdata
G121.30 & 1.2 & 2.07 & 3.20 &    0.4 &    0.2 & 0.42 \\
G123.07 & 2.2 & 3.66 & 4.47 &   11.5 &    3.5 & 0.30 \\
W32 & 2.4 & 5.50 & 6.02 &  490.0 &  124.0 & 0.25 \\
W3OH & 2.4 & 4.49 & 5.40 &   62.7 &   30.3 & 0.48 \\
GL490 & 0.9 & 3.01 & 3.53 &    3.1 &    0.4 & 0.13 \\
S231 & 2.3 & 3.33 & 4.04 &    5.9 &    1.3 & 0.22 \\
S231 & 1.6 & 3.72 & 4.40 &   13.0 &    3.0 & 0.23 \\
S241 & 4.7 & 3.79 & 4.72 &   15.2 &    6.2 & 0.41 \\
MonR2 & 0.9 & 4.27 & 4.74 &   40.3 &    6.5 & 0.16 \\
S252A & 1.5 & 3.29 & 4.18 &    5.5 &    1.8 & 0.33 \\
S255 & 1.3 & 3.79 & 4.56 &   15.2 &    4.4 & 0.29 \\
RCW142 & 2.0 & 4.26 & 5.04 &   38.9 &   13.2 & 0.34 \\
W28A2 & 2.6 & 5.17 & 5.85 &  251.9 &   83.7 & 0.33 \\
M8E & 1.8 & 4.30 & 4.93 &   42.9 &   10.2 & 0.24 \\
G9.62 & 7.0 & 4.92 & 5.82 &  152.2 &   79.4 & 0.52 \\
G8.67 & 4.5 & 4.06 & 4.97 &   26.2 &   11.2 & 0.43 \\
W311 &12.0 & 5.28 & 6.42 &  311.6 &  317.8 & 1.02 \\
G10.60 & 6.5 & 5.32 & 6.35 &  341.4 &  265.7 & 0.78 \\
G12.42 & 2.1 & 3.64 & 4.23 &   11.2 &    2.0 & 0.18 \\
G12.89 & 3.5 & 3.72 & 4.89 &   13.0 &    9.2 & 0.71 \\
G12.21 &13.7 & 5.36 & 6.40 &  371.6 &  302.4 & 0.81 \\
G13.87 & 4.4 & 4.77 & 5.41 &  111.3 &   30.7 & 0.28 \\
W33A & 4.5 & 4.79 & 5.57 &  115.1 &   44.4 & 0.39 \\
G14.33 & 2.6 & 3.32 & 4.57 &    5.8 &    4.4 & 0.76 \\
G19.61 & 4.0 & 4.79 & 5.60 &  115.4 &   47.7 & 0.41 \\
G20.08 & 3.4 & 3.93 & 4.87 &   20.0 &    8.8 & 0.44 \\
G23.95 & 5.8 & 4.91 & 5.63 &  148.7 &   50.8 & 0.34 \\
G24.49 & 3.5 & 4.26 & 5.25 &   39.6 &   21.3 & 0.54 \\
W42 & 9.1 & 5.97 & 6.74 & 1287.2 &  652.6 & 0.51 \\
G28.86 & 8.5 & 4.82 & 5.82 &  122.5 &   79.3 & 0.65 \\
W43S & 8.5 & 6.09 & 6.80 & 1649.3 &  760.7 & 0.46 \\
G31.41 & 7.9 & 4.30 & 5.40 &   43.0 &   29.8 & 0.69 \\
G31.44 &10.7 & 5.13 & 5.79 &  231.3 &   73.5 & 0.32 \\
W44 & 3.7 & 5.12 & 5.87 &  226.6 &   88.3 & 0.39 \\
S76E & 2.1 & 4.41 & 5.12 &   53.1 &   15.7 & 0.30 \\
G35.58 & 3.5 & 4.56 & 5.48 &   72.0 &   36.3 & 0.50 \\
G35.20 & 3.3 & 4.28 & 5.08 &   40.9 &   14.5 & 0.35 \\
W49 &14.0 & 6.56 & 7.33 & 4301.0 & 2530.4 & 0.59 \\
OH43.80 & 2.7 & 3.60 & 4.50 &   10.2 &    3.8 & 0.37 \\
G45.07 & 9.7 & 5.91 & 6.49 & 1133.5 &  366.5 & 0.32 \\
G48.61 &11.8 & 5.72 & 6.58 &  764.1 &  459.9 & 0.60 \\
W51W & 7.0 & 5.97 & 6.68 & 1280.4 &  567.7 & 0.44 \\
W51M & 7.0 & 6.58 & 7.15 & 4461.3 & 1695.4 & 0.38 \\
G59.78 & 2.2 & 3.52 & 4.31 &    8.7 &    2.4 & 0.28 \\
S87 & 1.9 & 4.13 & 4.77 &   30.4 &    7.0 & 0.23 \\
S88B & 2.1 & 4.54 & 5.24 &   69.9 &   20.6 & 0.29 \\
K350 & 9.0 & 5.94 & 6.59 & 1196.5 &  463.2 & 0.39 \\
ON1 & 6.0 & 4.01 & 5.20 &   23.6 &   19.0 & 0.81 \\
ON2 & 5.5 & 5.61 & 6.25 &  610.2 &  211.0 & 0.35 \\
S106 & 4.1 & 5.42 & 5.96 &  415.6 &  108.6 & 0.26 \\
G97.53 & 7.9 & 4.56 & 5.24 &   72.8 &   21.0 & 0.29 \\
BFS11B & 2.0 & 3.46 & 4.25 &    7.7 &    2.1 & 0.28 \\
CepA & 0.7 & 3.33 & 4.32 &    5.9 &    2.5 & 0.42 \\
S158 & 2.8 & 5.22 & 5.77 &  275.8 &   70.1 & 0.25 \\
NGC7538 & 2.8 & 5.21 & 5.77 &  272.8 &   70.8 & 0.26 \\
S157 & 2.5 & 4.16 & 4.89 &   31.8 &    9.2 & 0.29 \\
\enddata
\end{deluxetable*}

\begin{deluxetable*}{lrcrrrrrrrrrrr}
\tabletypesize{\scriptsize}
\tablecaption{Massive Dense Clump/Radio Continuum sample \label{tbl-5}}
\tablewidth{0pt}
\tablehead{
 \colhead{Object} & \colhead{Distance} & \colhead{Radio frequency} & \colhead{FWHM} & \colhead{Radio Flux$^1$} 
 & \colhead{Log(\lir)$^2$} & \colhead{SFR(radio)} & \colhead{\sfrlir} & \colhead{SFR(radio)/\sfrlir} \\
 \colhead{-} & \colhead{(kpc)} & \colhead{(GHz)} & \colhead{(arcmin)}  & \colhead{(Jy)} & \colhead{(\lsun)} &
\colhead{(\msun\ Myr$^{-1}$ )} & \colhead{(\msun\ Myr$^{-1}$ )} & \colhead{-}  \\
 }
\startdata
W28A2(1)	 & 2.6 & 4.875 & 4.0 & 5.5 & 5.71 & 58.0 & 62.0 & 0.94 \\
G9.62+0.19& 7.0 & \textquotedbl  & 2.8 & 1.12 & 5.66 & 41.9 & 54.1 & 0.77 \\
W31(1)& 12.0 & \textquotedbl & 3.1 & 1.11 & 6.21 & 150 & 196 & 0.77 \\
G10.60-0.40 & 6.5 & \textquotedbl & 3.2 & 4.46 & 6.14 & 188 & 167 & 1.1 \\
G12.21-0.10 & 13.7 & \textquotedbl & 3.3 & 1.67 & 6.23 & 333 & 205 & 1.6 \\
G13.87+0.28 & 4.4 & \textquotedbl & 2.7 & 3.83 & 5.31 & 52.7 & 24.3 & 2.2 \\
G19.61-0.23 & 4.0 & \textquotedbl & 2.9 & 4.98 & 5.33 & 65.3 & 25.7 & 2.5 \\
G20.08-0.13 & 3.4 & \textquotedbl & 3.0 & 1.13 & 4.61 & 11.5 & 4.83 & 2.4 \\
G23.95+0.16 & 5.8 & \textquotedbl & 2.7 & 2.32 & 5.38 & 55.5 & 28.4 & 2.0 \\
G24.49-0.04 & 3.5 & \textquotedbl & 3.1 & 0.62 & 4.64 & 7.12 & 5.26 & 1.4 \\
W43S & 8.5 & \textquotedbl & 5.0 & 4.5 & 6.64 & 792 & 527 & 1.5 \\
G31.41+0.31 & 7.9 & \textquotedbl & 2.6 & 1.2 & 5.04 & 49.4 & 12.6 & 3.7 \\
G31.44-0.26 & 10.7 & \textquotedbl & 3.4 & 1.06 & 5.77 & 137 & 70.6 & 1.9 \\
W44	& 3.7	 & \textquotedbl & 2.8 & 11.58 & 5.73 & 121 & 64.9 & 1.9 \\
G35.58-0.03 & 3.5 & \textquotedbl & 3.6 & 1.68 & 5.06 & 26.0 & 13.6 & 1.9 \\
G48.61+0.02 & 11.8 & \textquotedbl & 3.9 & 3.51 & 6.40 & 725 & 303 & 2.4 \\
W51W & 7.0 & \textquotedbl & 3.5 & 13.5 & 6.43 & 790 & 325 & 2.4 \\
W51M & 7.0 & \textquotedbl & 3.5 & 58 & 6.82 & 3390 & 783 & 4.3 \\
S76E & 2.1 & 5.000 & 9.5 & 7 & 5.20 & 20.4 & 18.9 & 1.1 \\
S87 & 1.9	& 2.695 & 9.0 & 2 & 4.86 & 4.48 & 8.67 & 0.52 \\
S88B & 2.1 & \textquotedbl & 7.0 & 9 & 5.30 & 24.7 & 23.7 & 1.0 \\
K3-50 & 9.0 & \textquotedbl & 8.5 & 23 & 6.67 & 1160 & 558 & 2.1 \\
\enddata

\tablenotetext{1}{This column gives a peak flux for 4.875 GHz data from A79 and an integrated
flux for the last four objects from A70.}
\tablenotetext{2}{\lir\ data obtained from photometry of IRAS images with aperture radius = FWHM.}
\end{deluxetable*}


\clearpage

\end{document}